\documentclass[a4paper,12pt]{article}

\usepackage{epsf,amsmath}
\usepackage{graphicx}

\usepackage{geometry}
\newcommand{\ice}[1]{\relax}

\newcommand{\ms}{\overline{\text{MS}}}

\begin{document}

\title{Non-Perturbative Approach to the Landau Gauge Gluodynamics}

\author{ A.Y.~Lokhov$^a$, Ph.~Boucaud$^b$, J.P.~Leroy$^b$, A.~Le~Yaouanc$^b$, \\
J. Micheli$^b$, O. P\`ene$^b$, J.~Rodr\'iguez-Quintero$^c$ and 
C.~Roiesnel$^a$ }
\date{ }
\maketitle
\begin{center}
$^a$ Centre de Physique Th\'eorique\footnote{
Unit\'e Mixte de Recherche 7644 du Centre National de 
la Recherche Scientifique}de l'Ecole Polytechnique\\
F91128 Palaiseau cedex, France\\ 
$^b$Laboratoire de Physique Th\'eorique et Hautes
Energies\footnote{Unit\'e Mixte de Recherche 8627 du Centre National de 
la Recherche Scientifique}\\
{Universit\'e de Paris XI, B\^atiment 211, 91405 Orsay Cedex,
France}\\
$^c$ Dpto. F\'isica Aplicada, Fac. Ciencias Experimentales,\\
Universidad de Huelva, 21071 Huelva, Spain.
\end{center}

\begin{abstract}
\noindent We discuss a non-perturbative lattice calculation of the ghost and gluon 
propagators in the pure Yang-Mills theory in Landau gauge. The ultraviolet behaviour
is checked up to NNNLO yielding the value $\Lambda^{n_f=0}_{\ms}=269(5)^{+12}_{-9}\text{ MeV}$, and
we show that lattice Green functions satisfy the complete Schwinger-Dyson equation for the 
ghost propagator for all considered momenta. The study of the above propagators at small momenta showed that the infrared 
divergence of the ghost propagator is enhanced, whereas the gluon propagator seem to remain 
finite and non-zero. The result for the ghost propagator is consistent with the 
analysis of the Slavnov-Taylor identity, whereas, according to this analysis, the gluon propagator
should diverge in the infrared, a result at odds with other approaches.
\end{abstract}
\begin{flushright}
\tt CPHT-PC076.1205, LPT-06-02, UHU-FT-18
\end{flushright}
%

%
\section{Introduction} 
%

In this talk we report on a non-perturbative lattice study of basic correlation functions
of the Euclidean Landau gauge pure Yang-Mills gauge theories, i.e. the gluon propagator
\begin{equation}
G^{(2)ab}_{\mu\nu}(p,-p) =
			\delta^{ab}\left(  \delta_{\mu\nu} - \frac{p_\mu p_\nu}{p^2}  \right)
			\frac{G(p)}{p^2},
\end{equation}
and the ghost propagator
\begin{equation}
F^{(2)ab}(p,-p) = \delta^{ab}\frac{F(p)}{p^2}.
\end{equation}
These propagators have been successfully checked by perturbation theory up to 
NNNLO in the ultraviolet domain,(\cite{Boucaud1},\cite{Boucaud2}) yielding the non-perturbative 
value of
\begin{equation}
\Lambda^{n_f=0}_{\ms}=269(5)^{+12}_{-9}\text{ MeV}.
\end{equation}
Now we concentrate  on  the infrared exponents $\alpha_F$ and $\alpha_G$
that describe power-law deviations from free propagators  when $p\rightarrow 0$
\begin{equation}
G(p^2)\propto (p^2)^{\alpha_G} \qquad F(p^2)\propto (p^2)^{\alpha_F}.
\end{equation}
%

%
\section{Non-perturbative study of the infrared exponents $\alpha_F,\alpha_G$}
%

Diverse analytical approaches (study of truncated Schwinger-Dyson equations
and of renormalisation group equation~\cite{LercheZwanzigerFischer},~\cite{Bloch})
agree that the infrared divergence of the ghost propagator is enhanced,
i.e. $\alpha_F<0$; while they predict different values for $\alpha_G$,
mostly around $\alpha_G\approx 1.2$. This means that the gluon
propagator is suppressed in the infrared. Lattice simulations confirmed the
prediction for the ghost propagator, whereas the lattice gluon propagator
seems to remain finite and non-zero in the infrared, i.e. $\alpha_G = 1$~\cite{Gluon_Finie_a_zero}.

Let us consider the Slavnov-Taylor identity~\cite{ST} relating the three-gluon vertex
$\Gamma_{\lambda \mu \nu}$, the ghost-gluon vertex $\widetilde{\Gamma}_{\lambda\mu}(p,q;r)$
\footnote{$r$ is the momentum of the gluon, $q$ is the momentum of the entering ghost.}
and the propagators:
\begin{equation}
\label{ST}
p^\lambda\Gamma_{\lambda \mu \nu} (p, q, r)  =
\frac{F(p^2)}{G(r^2)} (\delta_{\lambda\nu} r^2 - r_\lambda r_\nu) \widetilde{\Gamma}_{\lambda\mu}(r,p;q) -
\frac{F(p^2)}{G(q^2)} (\delta_{\lambda\mu} q^2 - q_\lambda q_\mu) \widetilde{\Gamma}_{\lambda\nu}(q,p;r).
\end{equation}
Taking the limit $r \rightarrow 0$ keeping $q$ and $p$ finite, and
using $G(r^2) \simeq \left( r^2\right)^{\alpha_G}$, one finds the following
limits on the infrared exponents~\cite{SDST}
\begin{equation}
\label{STpredictions}
\left\{
\begin{array}{ll}
\alpha_G < 1  & \text{gluon propagator \emph{diverges} in the infrared, and} \\
\alpha_F \le 0 & \text{the divergence of the ghost propagator is \emph{enhanced} in the infrared}.
\end{array}
\right.
\end{equation}
The limit on $\alpha_G$ disagrees with many other analytical predictions
~\cite{LercheZwanzigerFischer}, except some cases in~\cite{Bloch}.
Let us try to understand this discrepancy. All these methods rely on a commonly 
accepted relation between the infrared exponents
\begin{equation}
\label{R}
2\alpha_F + \alpha_G = 0.
\end{equation}
which we shall discus now. The origin of this relation is the dimensional analysis of the Schwinger-Dyson equation for the 
ghost propagator(SD):
\begin{equation}
\label{SDghost}
\frac{1}{F(k)}  = 1 + g_0^2 N_c \int \frac{d^4 q}{(2\pi)^4} 
\left( \rule[0cm]{0cm}{0.8cm}
\frac{F(q^2)G((q-k)^2)}{q^2 (q-k)^2} 
\left[  
\frac{(k\cdot q)^2 - k^2q^2}{k^2(q-k)^2}  
\right]
\  H_1(q,k) 
\right),
\end{equation}
where $H_1(q,k)$ is one of the scalar functions defining the ghost-gluon vertex:
\begin{equation}
q_{\nu'} \widetilde{\Gamma}_{\nu'\nu}(-q,k;q-k) =  q_\nu H_1(q,k) + (q-k)_\nu H_2(q,k).
\end{equation}
The large momentum behaviour(\cite{Chetyrkin:2000dq},\cite{ST}) of this vertex depends on the kinematic configuration:
\begin{equation}
\begin{array}{l}
\frac{p_{\mu}p_{\nu}}{p^2} \cdot \widetilde{\Gamma}_{\mu\nu}^{\overline{\text{MS}}}(-p,0;p) = 1\quad\text{to \emph{all} orders}
\\
\frac{p_{\mu}p_{\nu}}{p^2} \cdot \widetilde{\Gamma}_{\mu\nu}^{\overline{\text{MS}}}(-p,p;0) = 1 + \frac{9}{16\pi}\alpha_s^2(p^2) + \ldots
\end{array}
\end{equation}
Note that in the case of the vanishing momentum of the out-going ghost (and only in this case) 
one has
\begin{equation}
H_1(q,0) + H_2(q,0)=1.
\end{equation}
If \emph{both} $H_{1,2}$ are non-singular then one can suppose $H_1(q,k)\simeq 1$ in (\ref{SDghost}),
and (\ref{R}) is straightforward by a dimensional analysis. However, we have a priori no reason
to think that the scalar functions $H_1(q,k)$ and $H_2(q,k)$ are \emph{separately} non-singular for all
$q,k$. Writing, for example, $H_1(q,k) \ \sim \ (q^2)^{ \alpha_\Gamma } \ h_1\left(\frac{q\cdot k}{q^2},\frac{k^2}{q^2} \right) $,
with a non-singular function $h_1$, we keep all the generality of the argument admitting a singular behaviour of
the scalar factor $H_1(q,k)$. Doing the dimensional analysis of eq. (\ref{SDghost}) \emph{without}
putting $H_1(q,k)\simeq 1$, we obtain that the relation (\ref{R}) is satisfied
if and only if the following triple condition is verified~\cite{SDST}:
\begin{equation}
\label{conditions}
2\alpha_F + \alpha_G = 0 \quad \Longleftrightarrow \quad
\left\{ 
\begin{array}{l}
\alpha_F \ne 0 \\
\alpha_\Gamma=0 \\
\alpha_F +\alpha_G < 1
\end{array}
\right.
\end{equation}
The first and the last conditions are compatible with limits coming from the analysis 
of the Slavnon-Taylor identity (\ref{STpredictions}), and are also consistent with 
our lattice simulations~\cite{SDST}. If one of the conditions (\ref{conditions}) is not 
verified then (\ref{R}) should be replaced by
\begin{equation}
\label{R_Gamma}
2\alpha_F + \alpha_G + \alpha_\Gamma = 0.
\end{equation}
In the following section we present the results of a numerical test of the relation (\ref{R}),
and thus we probe the validity of the condition on $\alpha_\Gamma$.

%
\section{Lattice check of analytical predictions}
%

\begin{figure}[!thb]
	\begin{center}
	\includegraphics[angle=-90, width=0.9\linewidth]{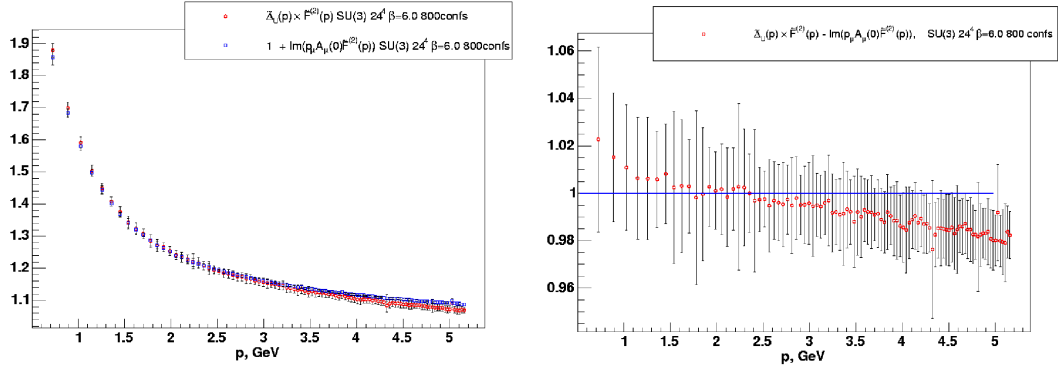}
	\end{center}
\caption[*]{\footnotesize Checking that lattice Green functions satisfy the ghost SD equation (\ref{SDlattice}). 
The l.h.s vs r.h.s of (\ref{SDlattice}) is plotted at the left, and at the right we plot 
$\widetilde{F}(p^2) -  g_0 \frac{p_\mu}{N_c^2 -1} f^{abc}\langle A^c_\mu(0).
\widetilde{F}_{\text{1conf}}^{(2)ba}(\mathcal{A},p)\rangle $ compared to $1$}
\label{SDcheck}
\end{figure}

\begin{figure}[!thb]
\begin{center}
	\includegraphics[angle=-90, width=0.6\linewidth]{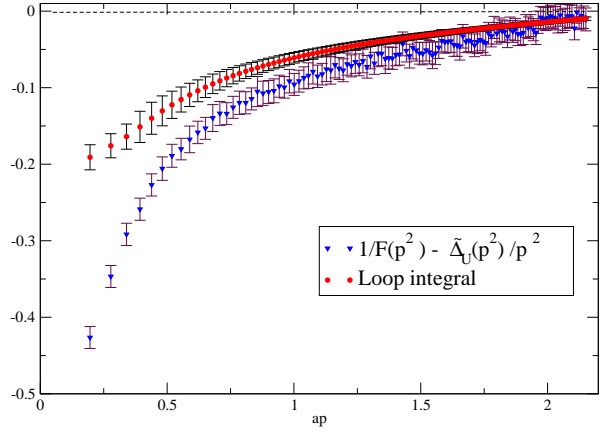}  
\end{center}
\caption[*]{\footnotesize Checking whether lattice Green functions satisfy the ghost SD equation (\ref{SDghost}) with
an assumption $H_1(q,k)=1$. The upper line(circles) correspond to the loop integral in (\ref{SDghost}), 
and the down line (triangles) corresponds to $1/F(p^2)-1$. In this plot $a^{-1}\approx 3.6$ GeV. }
\label{SDcheckH1}
\end{figure}

As we have already mentioned, lattice Green functions are checked at high precision
with perturbation theory at large momentum. In order to test the validity of 
lattice predictions at small momenta we verified that lattice Green functions
satisfy the ghost SD equation written in the form
\begin{eqnarray}
\label{SDlattice}
\widetilde{F}(p^2)=  1 + g_0 \frac{p_\mu}{N_c^2 -1} f^{abc}\langle A^c_\mu(0)\cdot
\widetilde{F}_{\text{1conf}}^{(2)ba}(\mathcal{A},p)\rangle ,
\end{eqnarray}
where $\widetilde{F}_{\text{1conf}}^{(2)ba}(\mathcal{A},p)$ is the correlator of the 
ghost and anti-ghost fields in a background field $\mathcal{A}$.
Note that the three-point function in the r.h.s is in a mixed coordinate-momentum 
representation, and that here we make \emph{no assumption} about the vertex. 
The traditional form (\ref{SDghost}) differs from (\ref{SDlattice}) by a 
Legendre transformation. We see from Fig.\ref{SDcheck} that the lattice Green functions 
match pretty well the SD equation (\ref{SDlattice}) in both the ultraviolet and
infrared regions. 
%
%
Now we try to check the validity of the non-singularity assumption for the scalar
factor $H_1(q,k)$. For this purpose we check numerically whether our lattice propagators
satisfy the equation (\ref{SDghost}) completed with the assumption $H_1(q, k ) = 1$ 
(cf. Fig.\ref{SDcheckH1}~\cite{SDST}). We see that at small momenta (below $\approx3$ GeV) 
the SD \emph{with the assumption} $H_1(q,k)=1$ is not satisfied. This suggests that
%
%
the scalar function $H_1(q,k)$ plays an important role in the infrared gluodynamics.
%
\begin{figure}[!thb]
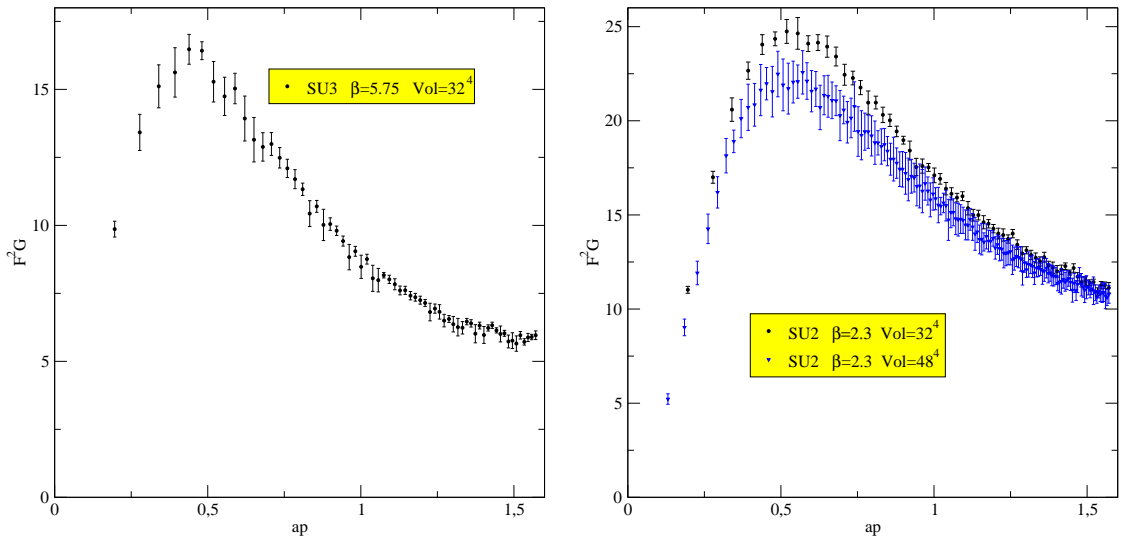

\begin{center}
	\begin{tabular}{lr}
	\includegraphics[width=0.45\linewidth]{SU3_F2G.eps}&\includegraphics[width=0.45\linewidth]{SU2_F2G.eps}
	\end{tabular}
\end{center}
\caption[*]{\footnotesize Direct test of the relation $2\alpha_F + \alpha_G=0$. If the last is true $F^2G$ has to be
constant in the infrared. We see that it is clearly not the case. In these plots  $a^{-1}\approx 1.2$ GeV, so the peak is located at $\approx 600$ MeV.}
\label{F2G}
\end{figure}
%
Finally, a direct test of the relation (\ref{R}) and thus, as we have seen
in (\ref{conditions}), of the value of $\alpha_\Gamma$ can be done. For this
we plot the quantity $F^2(p^2) G(p^2)$~Fig.\ref{F2G}. 
If all the conditions (\ref{conditions}) are satisfied this quantity should be constant 
in the infrared. We see from Fig.\ref{F2G} that in the infrared (below $\approx 600$ MeV) 
the quantity $F^2 G$ is not constant, and thus one of the conditions (\ref{conditions})
is not verified. We have seen that the conditions $\alpha_F \neq 0 $ and $\alpha_F+\alpha_G < 1$
are consistent with the limits (\ref{STpredictions}) from the Slavnov-Taylor
identity (\ref{ST}). We have also seen (cf. Fig.\ref{SDcheck},\ref{SDcheckH1})
that neglecting the vertex is not possible in the infrared, because in this case
the ghost SD equation is no longer satisfied by lattice propagators.
Thus the only possibility is to admit that $H_1(q,k)$ plays an important role,
and that the relation (\ref{R}) is not verified.  The modified form (\ref{R_Gamma})
that takes in account the singularity of $H_1(q,k)$ should be
considered (according to Fig.\ref{SDcheckH1}), with $\alpha_\Gamma < 0$
in our parametrisation.

%
\section{Conclusions}
%

We have seen that lattice simulations allow a whole momentum range study of
Green functions of a non-Abelian gauge theory in Landau gauge. They have been tested by
perturbation theory up to NNNLO at large momentum. We have also checked
(numerically) that lattice Green functions satisfy the complete ghost Schwinger-Dyson equation
(\ref{SDlattice}) for all considered momenta. These tests allow us to conclude that
numerical simulation on the lattice give relevant results not only
in the ultraviolet domain but also in the infrared one.

Our analysis of the Slavnov-Taylor identity showed that the power-law
infrared divergence of the ghost propagator is enhanced in the infrared
(compared to the free case), and that the gluon propagator must diverge in the infrared.
The latter limit is in conflict with most present
analytical estimations~\cite{LercheZwanzigerFischer}, that
support a vanishing gluon propagator in the infrared. In~\cite{Bloch} the author
replaced $H_1(q,k)$ by its perturbative expansion, and an interval of
values of $\alpha_G$ was found, including those compatible with (\ref{ST}).
Lattice simulations point to a finite non-vanishing infrared gluon propagator,
in conflict with the above analysis of the Slavnov-Taylor identity.

Our numerical studies showed that the commonly accepted relation (\ref{R}) between the infrared exponents
is not valid, because $F^2 G$ is infrared suppressed, and hence $2\alpha_F + \alpha_G > 0$.
This statement is supported by the fact that lattice propagators do not match the reduced
SD equation (Fig.\ref{SDcheckH1}), whereas the complete one is perfectly
verified (Fig.\ref{SDcheck}). This speaks in favour of a singularity in the scalar factor $H_1$.
Note that it does not contradict the non-renormalisation theorem~\cite{ST} which implies
for the renormalisation constant of the ghost-gluon vertex :
$\widetilde{Z}^{\overline{\text{MS}}}_1 = 1$.

\end{document}